\documentclass[10pt,journal,compsoc]{IEEEtran}

\ifCLASSOPTIONcompsoc
\usepackage[nocompress]{cite}
\else
\usepackage{cite}
\fi

\usepackage[table]{xcolor}

\usepackage{amsmath}
\allowdisplaybreaks
\usepackage{amssymb}
\usepackage{amsthm}

\usepackage[ruled,vlined]{algorithm2e}

\usepackage{graphicx}
\usepackage{float}
\usepackage{caption}
\usepackage{multirow}

\newtheorem{lemma}{Lemma}
\newtheorem{claim}{Claim}
\newtheorem{remark}{Remark}


\usepackage{booktabs}
\usepackage{array}

\usepackage[hidelinks]{hyperref}

\title{BiLight: A Fast Method for Estimating Counts of 6-Vertex Bipartite Motifs in Large Graphs}

\author{Mustafa~\c{C}elik,~Özgür~Okumuş,~and~Lale~Özkahya%
\IEEEcompsocitemizethanks{\IEEEcompsocthanksitem M.~\c{C}elik and Ö.~Okumuş are with the
Department of Computer Engineering, and L.~Özkahya is with the Department of Artificial
Intelligence and Data Engineering, Hacettepe University, Ankara, Turkey.}%
}

\IEEEtitleabstractindextext{%
    \begin{abstract}
        Motifs, or small subgraphs, are fundamental building blocks of complex networks. Analyzing their frequency distribution provides deep insights into the local structure of graphs, with applications in social network analysis, bioinformatics, and web mining. However, exact enumeration of motifs becomes computationally expensive as the size of the motif and the network increases. This challenge is particularly acute for bipartite motifs, which are ubiquitous in modeling relationships like user-item, author-paper, and actor-movie interactions.
        In this work, we present \textbf{BiLight} (\textbf{Bi}partite \textbf{Light}weight motif estimation), a scalable framework for estimating the frequencies of all 17 non-isomorphic 6-vertex bipartite motifs. Specifically, we focus on estimating the non-induced counts of these motifs, which is a common requirement in many graph mining tasks.
        Our approach relies on a randomized path sampling of paths on four vertices. We use an efficient, weighted strategy by counting the ways that each sampled path can be extended to one of the 17 motifs. This framework  yields unbiased estimates of the global motif counts. We conduct extensive experiments on networks of various sizes.
        Our experiments demonstrate that BiLight converges rapidly to the ground truth with high accuracy, even on massive networks where exact counting is infeasible.
        The source code is publicly available at \href{https://github.com/celikmustaa/BiLight}{\nolinkurl{github.com/celikmustaa/BiLight}}.
    \end{abstract}

    \begin{IEEEkeywords}
        Graphlets, motif counting, bipartite graphs, path sampling, randomized algorithms, BiLight.
    \end{IEEEkeywords}
}

\begin{document}

    \maketitle
    \IEEEdisplaynontitleabstractindextext
    \IEEEpeerreviewmaketitle

    \IEEEraisesectionheading{\section{Introduction}\label{sec:introduction}}

\IEEEPARstart{M}{otifs} are small subgraphs on networks and
counting how often they occur is central to network
analysis, with uses in biology, the social sciences and the study of
large online networks
\cite{holland1970method, coleman1988social, portes1998social, milo2002network, burt2004structural, przulj2004modeling, fagiolo2007clustering, hormozdiari2007scale, becchetti2008efficient, faust2010puzzle, szell2010measuring, welles2010friend, son2012analysis}.
Motif frequency in real networks is an important feature
in comparison to a random network~\cite{holland1970method, watts1998collective, milo2002network}.
The frequency distribution of motifs, often called the \textit{motif degree distribution} or \textit{motif profile}, provides a powerful signature of the network's topology.
Motif profile is also used as a graph feature
to evaluate the performance of graph models representing networks
and classify vertex roles~\cite{holland1970method, przulj2004modeling, burt2004structural, hormozdiari2007scale, fagiolo2007clustering, becchetti2008efficient, seshadhri2012community, durak2012degree, ugander2013subgraph}.
Understanding the local structure of these networks is crucial for tasks like community detection, link prediction, anomaly detection, and graph similarities for graph classification, which have been
widely used for a variety of graph mining and learning tasks.

Milo et al.~\cite{milo2002network} introduced network motifs as
subgraphs that occur in real world networks more frequently when
compared to random networks. Network motifs have been significant
structures in the study of transcriptional regulation networks~\cite{moss5_ref02_shenorr2002}, the analysis of online social networks~\cite{moss5_ref03_chun2008cyworld, moss5_ref04_slashdot2009, moss5_ref05_douban2011, ugander2013subgraph} and internet traffic classification and anomaly detection~\cite{moss5_ref07_jin2009traffic, moss5_ref08_iliofotou2009dynamicity}.

Graphlet (motif as induced subgraphs) counts are useful
as significant graph features in graph classification and
graph similarity comparison.
Two widely used measures are the graphlet kernel~\cite{moss5_ref09_shervashidze2009graphlet},
which scores a pair of graphs by the inner product of their normalized
graphlet-count vectors, and the relative graphlet frequency (RGF)
distance~\cite{przulj2004modeling}, which sums the absolute differences of the log-scaled relative frequencies.
Such comparisons are needed in computational biology,
cheminformatics and software analysis, among other fields. For example,
protein function prediction problem asks whether a given protein
is an enzyme. The topology of a biological network of a protein is
used to determine this characteristic of a protein
~\cite{moss5_ref11_dobson2003enzymes, moss5_ref12_vishwanathan2010kernels}.
Similarly, the networks representing chemical compounds are used
for predicting activity
or adverse effects of potential drugs~\cite{moss5_ref13_wale2006descriptors, yanardag2015smoothing}.
Computing similarities of networks using graph models is also useful
in community detection and clustering problems.
For example,
Yanardag and Vishwanathan~\cite{moss5_ref15_yanardag2015deep}
compute graph similarity to determine the type of a community
in online social networks.
Further examples on application of graph similarity comparison
arise in malware detection, where subroutines and classes in softwares
and relationships are represented with a graph model~\cite{moss5_ref16_chen2015malice, moss5_ref17_shen2014android, moss5_ref18_gascon2013android} so that similarities between
the topologies of these graphs help to identify the malwares.

\begin{figure}[htbp]
        \centering
        \includegraphics[width=0.90\linewidth]{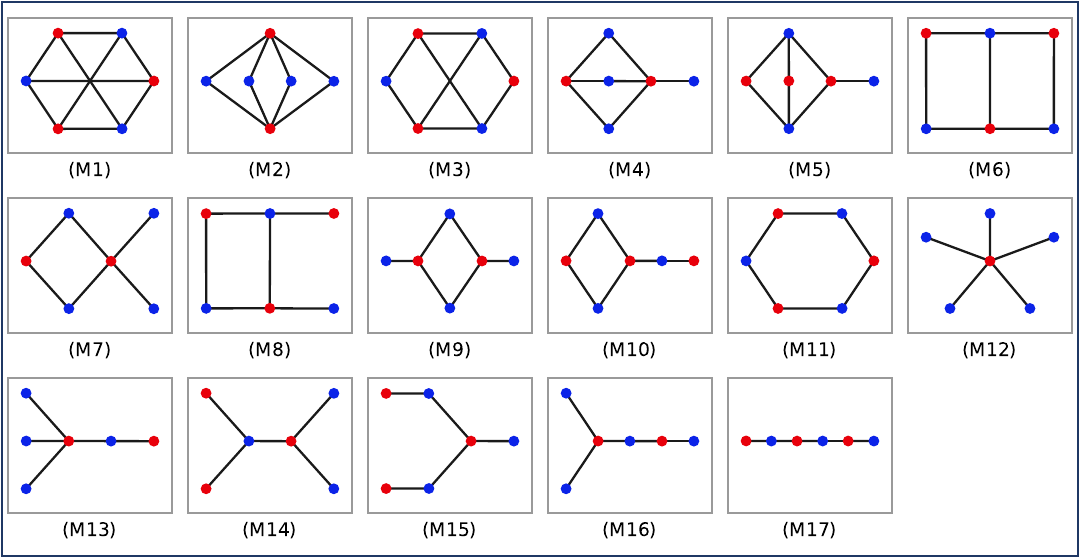}
        \caption{The 17 non-isomorphic 6-vertex bipartite motifs (M1--M17).}
        \label{fig:motifs}
    \end{figure}

Motif analysis is used widely in graph mining problems~\cite{sariyuce2018peeling,seshadhri2013triadic,ahmed2017graphlet}.
In these applications, the computation of the motif counts
is an expensive process even for moderate size graphs
due to the combinatorial explosion of the problem~\cite{betzler2011parameterized, wong2012biological, rahman2014graft}.
Despite the progress on counting 4-vertex motifs~\cite{ahmed2015efficient, jha2015path} and 5-vertex motifs~\cite{pinar2017escape, wang2017moss}, little attention has been given to counting 6-vertex motifs: existing work targets a single pattern, such as the induced 6-cycle~\cite{niu2022counting, niu2025fast}, or general non-bipartite graphlets~\cite{bressan2018motif}.
While counting 3-vertex or 4-vertex motifs (like triangles in unipartite graphs or butterflies in bipartite graphs) is well-studied, counting larger motifs (e.g., 6 vertices) presents a significant computational challenge. The number of such subgraphs grows combinatorially with the graph size, making exact enumeration impossible for large-scale networks. Consequently, researchers have turned to sampling-based approximation methods~\cite{wang2017moss, bressan2018motif}.
In this work, we present BiLight (Bipartite Lightweight motif estimation), a scalable framework for estimating the non-induced counts of all 17 non-isomorphic 6-vertex bipartite motifs shown in Figure~\ref{fig:motifs}. While these motifs are bipartite, the input graph $G$ need not be; BiLight correctly counts these bipartite substructures within any general graph.

 Our contribution can be summarized as follows:
 \begin{itemize}
     \item BiLight's core contribution is a method to sample paths on four vertices uniformly at random and extend them to estimate 6-vertex bipartite motifs counts. Our method is scalable and computationally efficient.
    \item To validate our method, we conduct extensive experiments
    to show the accuracy of our method. We provide theoretical analysis to show that our framework finds unbiased estimates
of 6-vertex bipartite motif counts. To do that, we bound the errors of the estimated counts finding bounds on the variance of the estimator.
    \item We conduct experiments on real networks with various scales such as Amazon Ratings (Books), Amazon Ratings (Electronics), IMDB, and Flickr. Our experiments demonstrate that BiLight converges rapidly to the ground truth with high accuracy (the ground truth is approximated by high-sample runs), achieving low relative error and high stability even on massive graphs where exact counting is infeasible. Our empirical comparisons show that BiLight outperforms the state-of-the-art method.
    To allow the experiments to be repeated independently, we release the
source code of BiLight, together with the test suite that
checks the local counting patterns against a brute-force oracle.
 \end{itemize}
The rest of this paper is organized as follows.
In Section~\ref{sec:preliminaries}, we provide necessary notation
and terminology.
Section~\ref{sec:related} summarizes the related work.
The preprocessing step of our framework is described in Section~\ref{sec:preprocess}.
Section~\ref{sec:samplingAlgo} presents our 6-vertex bipartite motif sampling method BiLight.
The performance evaluation and testing results are presented
in Section~\ref{sec:results}.






    \section{Preliminaries}
    \label{sec:preliminaries}
    Let $G=(V,E)$ be an undirected graph with $n=|V|$ vertices and $m=|E|$ edges. For a vertex $v \in V$, let $\Gamma(v)$ denote its neighbor set and $d(v)=|\Gamma(v)|$ its degree. A graph $G$ is \textit{bipartite} if its vertex set can be partitioned into two disjoint sets $U$ and $W$ such that every edge connects a vertex in $U$ to a vertex in $W$.
    A \textit{motif} $M$ is a small connected graph used as a pattern.
    An \emph{induced subgraph} $G'=(V',E')$ is a subgraph of $G$ such that
    $E'$ is exactly set of edges in $E$ that have both its endvertices in $V'$. For example, the cycle $C_3$ has three copies of $P_3$ as a subgraph, but none of them are induced as the endvertices of these paths are connected with an edge in $C_3.$
    The \textit{non-induced count} $N(M)$ is the number of subgraphs of $G$ isomorphic to $M$ (additional edges among the chosen vertices are allowed), whereas the \textit{induced count} is the number of vertex subsets $S$ whose induced subgraph $G[S]$ is isomorphic to $M$ (no additional edges). A motif counted in the induced sense is commonly called a \textit{graphlet}.
    A path (cycle, resp.) on $n$ vertices is denoted by $P_n$ ($C_n$, resp.).
    Table~\ref{tab:notation} summarizes the notation used throughout the paper.

    \begin{table}[htbp]
        \centering
        \footnotesize
        \setlength{\tabcolsep}{3pt}
        \renewcommand{\arraystretch}{1.0}
        \caption{Notation used throughout the paper.}
        \label{tab:notation}
        \begin{tabular}{@{} c >{\raggedright\arraybackslash}p{\dimexpr\columnwidth-6.2em} @{}}
            \toprule
            \textbf{Symbol} & \textbf{Description} \\
            \midrule
            $G=(V,E)$ & Undirected graph with vertex set $V$ and edge set $E$ \\
            $n, m$ & Number of vertices $|V|$ and edges $|E|$ \\
            $\Gamma(v)$ & Neighbor set of vertex $v$ \\
            $d(v)$ & Degree of vertex $v$, i.e., $|\Gamma(v)|$ \\
            $\Delta$ & Maximum degree in $G$ \\
            $\alpha$ & Arboricity of $G$: the minimum number of forests into which $E$ can be partitioned \\
            $P_n$ & Path on $n$ vertices ($n-1$ edges) \\
            $C_n$ & An $n$-cycle (cycle on $n$ vertices) \\
            $N(P_4)$ & Non-induced count of $P_4$ in $G$ \\
            $M_i$ & The $i$-th bipartite 6-vertex motif ($i = 1, \ldots, 17$) \\
            $N(M)$ & Non-induced count of motif $M$ in $G$ \\
            $\widehat{N}(M)$ & Estimated non-induced count of motif $M$ in $G$ \\
            $c_M$ & Number of distinct copies of $P_4$ inside a single instance of motif $M$ \\
            $A, A^{-1}$ & Conversion matrices between induced and non-induced counts \\
            $k$ & Sample budget: the number of sampled $P_4$ \\
            $w_e$ & Edge weight of $e=(u,v)$ in weighted $P_4$ sampling: $(d(u)-1)(d(v)-1)-|\Gamma(u)\cap\Gamma(v)|$ \\
            $\operatorname{ext}(M,P)$ & Number of ways a sampled $P_4$ extends to motif $M$ in $G$ \\
            \bottomrule
        \end{tabular}
    \end{table}

\section{Related Work}\label{sec:related}
    In this section, we review various related works on counting motifs in networks.
    Subgraph and motif counting is a broad and well-studied area; we refer the reader to Ribeiro et al.~\cite{ribeiro2021survey} for a comprehensive survey.

    \textit{Exact counting.}
    Triangle counting has received the most attention, exactly~\cite{chiba1985arboricity} as well as by approximation~\cite{tsourakakis2011triangle, seshadhri2013triadic}.
    Gonen and Shavitt~\cite{gonen2009approximating}
give approximation algorithms for the non-induced counts of all
4-vertex motifs, including per-vertex counts, while Marcus and Shavitt~\cite{marcus2010efficient} count the full set of
4-vertex motifs exactly. More progress and faster algorithms on counting 4-vertex motifs are
developed by Ahmed et al.~\cite{ahmed2015efficient, ahmed2017graphlet} by utilizing the relationships
between 3-vertex and 4-vertex motif counts.
     P\i nar et al.~\cite{pinar2017escape} compute exact counts of all subgraphs on up to five vertices. Beyond five vertices, exact counting becomes provably hard: Bera et al.~\cite{bera2022counting} show that in bounded-degeneracy graphs (which include most real-world networks) no near-linear-time algorithm can count a pattern containing an induced cycle of length six or more, unless standard fine-grained complexity conjectures fail. This barrier at size six is a central motivation for \emph{estimating} rather than exactly counting 6-vertex motifs.

    \textit{Approximating Motif Counts.}
    There have been studies devoted to designing sampling methods
for computing motif concentrations~\cite{bhuiyan2012guise, rahman2014graft, moss5_ref39_wang2014motif, moss5_ref48_kashtan2004sampling, wernicke2006efficient, moss5_ref50_omidi2009moda, moss5_ref51_chen2016randomwalk}.
    BiLight belongs to a family of sampling-based estimators. Wedge sampling~\cite{schank2005approximating, kolda2014counting, seshadhri2013triadic} draws random 2-paths and tests for closure to estimate triangle counts
and clustering coefficients. Alon et al.~\cite{alon1994colorcoding} propose the method of color-coding that reduces the computation cost by
    coloring vertices randomly and enumerating only colorful motifs. In~\cite{jha2015path}, path sampling is reported to run hundreds of times
faster than color coding on graphs with about a million edges, at comparable accuracy.
    In~\cite{tsourakakis2009doulion, pavan2013counting, moss5_ref23_jha2013birthday, moss5_ref24_ahmed2014samplehold},
    sampling method is used to estimate
the number of triangles of static and dynamic graphs.
Jha et al.~\cite{jha2015path} generalize this to \emph{path sampling}, selecting random 3-paths ($P_4$) to estimate the counts of all 4-vertex subgraphs with provable error bounds. Wang et al.~\cite{wang2017moss}
design a sampling algorithm to estimate the counts of 5-vertex motifs.
    For larger patterns, Bressan et al.~\cite{bressan2018motif} use color coding to estimate induced counts of 5- to 8-vertex graphlets in general graphs; their tool Motivo~\cite{bressan2019motivoarxiv, bressan2019motivo, bressan2021faster} speeds this up with succinct data structures and adaptive sampling. BiLight extends the path-sampling paradigm one step further, sampling $P_4$ subgraphs and extending each to estimate all 6-vertex \emph{bipartite} motifs simultaneously.

    \section{Preprocessing}\label{sec:preprocess}
    BiLight estimates the non-induced count of each 6-vertex bipartite motif in two phases:
    \emph{preprocessing} (Algorithm~\ref{alg:preprocess}) builds an alias table for uniform $P_4$ sampling, and
    \emph{estimation} (Algorithm~\ref{alg:estimate}) draws $k$ random $P_4$ subgraphs and accumulates local extension counts.
    The \emph{seed} structure is a $P_4$, which is common to 16 of the 17 motifs (all except the star $M_{12}$).
    \begin{algorithm}[htbp]
    \DontPrintSemicolon
    \caption{BiLight: Preprocessing}\label{alg:preprocess}
    \KwIn{Graph $G=(V,E)$ with edges $E=\{e_1,\dots,e_m\}$}
    \KwOut{Arrays $\mathit{prob}[i]$, $\mathit{alias}[i]$ for $i=1,\dots,m$;\; total $P_4$ count $N(P_4)$}
    \ForEach{edge $e_i=(u,v) \in E$}{
      $w_{e_i} \gets (d(u)-1)\,(d(v)-1) - |\Gamma(u)\cap\Gamma(v)|$\tcp*{number of $P_4$ centered at $e_i$}
    }
    $N(P_4) \gets \sum_{i=1}^{m} w_{e_i}$\;
    \For{$i \gets 1$ \KwTo $m$}{
      $q_i \gets m\, w_{e_i} / N(P_4)$\tcp*{scaled by $m$ so the mean is $1$}
    }
    $\mathit{Small} \gets \{i : q_i < 1\}$;\quad $\mathit{Large} \gets \{i : q_i \ge 1\}$\;
    \While{$\mathit{Small}$ and $\mathit{Large}$ are both non-empty}{
      remove an index $s$ from $\mathit{Small}$ and an index $\ell$ from $\mathit{Large}$\;
      $\mathit{prob}[s] \gets q_s$;\quad $\mathit{alias}[s] \gets \ell$\;
      $q_\ell \gets q_\ell - (1 - q_s)$\tcp*{bin $\ell$ covers the shortfall of bin $s$}
      \eIf{$q_\ell < 1$}{move $\ell$ to $\mathit{Small}$}{move $\ell$ to $\mathit{Large}$}
    }
    \ForEach{index $i$ remaining in $\mathit{Small}$ or $\mathit{Large}$}{
      $\mathit{prob}[i] \gets 1$\;
    }
    \Return{$\mathit{prob}[\cdot],\ \mathit{alias}[\cdot],\ N(P_4)$}\;
    \end{algorithm}

    \begin{claim}\label{clm:uniformrandom}
       For any path $P$ on 4 vertices, the probability that $P$ is picked in Algorithm~\ref{alg:estimate} is $1/N(P_4)$.
    \end{claim}
    \begin{proof}

    Let $P$ be a path on distinct vertices $(v_0,v_1,v_2,v_3)$ with \emph{central edge} $e=\{v_1,v_2\}$ and \emph{extension vertices} $\{v_0, v_3\}$, and let
    \[
    w_e := (d(v_1)-1)(d(v_2)-1) - |\Gamma(v_1)\cap\Gamma(v_2)|,
    \]
    so that $N(P_4)=\sum_{e\in E} w_e$. Indeed, the product counts the ordered pairs $(v_0,v_3) \in (\Gamma(v_1)\setminus\{v_2\}) \times (\Gamma(v_2)\setminus\{v_1\})$, and the subtracted term removes the degenerate pairs with $v_0=v_3$: these are exactly the common neighbors of $v_1$ and $v_2$, and they close a triangle instead of forming a $P_4$. On a triangle-free host---in particular a bipartite one---this term vanishes and $w_e=(d(v_1)-1)(d(v_2)-1)$.

    To draw $P$ uniformly at random among all $P_4$'s, we apply the following steps: \emph{(i)}~select a central edge $e$ with probability $w_e/N(P_4)$; \emph{(ii)}~select $v_0$ uniformly at random from $\Gamma(v_1)\setminus\{v_2\}$ and $v_3$ uniformly at random from $\Gamma(v_2)\setminus\{v_1\}$; and \emph{(iii)}~if $v_0=v_3$, reject the pair and repeat step~\emph{(ii)}. Step~\emph{(ii)} draws each of the $(d(v_1)-1)(d(v_2)-1)$ ordered pairs with equal probability, so conditioning on acceptance leaves each of the $w_e$ valid pairs equally likely, that is, with probability $1/w_e$. These steps yield
    \[
    \Pr(P)= \frac{w_e}{N(P_4)} \cdot \frac{1}{w_e} = \frac{1}{N(P_4)}.
    \]
    Rejecting the whole pair (rather than redrawing a single endpoint) is what keeps the accepted path uniform. The step is also inexpensive. Write $a=d(v_1)-1$, $b=d(v_2)-1$ and $c=|\Gamma(v_1)\cap\Gamma(v_2)|$, so that $w_e=ab-c$ and the acceptance probability is $1-c/(ab)$. If $\max(a,b)\ge 2$ then $c\le\min(a,b)$ gives $c/(ab)\le 1/\max(a,b)\le 1/2$; and if $a=b=1$ then $w_e>0$ forces $c=0$, so the draw is always accepted. Hence every edge with $w_e>0$ is accepted with probability at least $1/2$, and the expected number of draws is at most two. In practice the step is far cheaper than this bound: across the twelve non-bipartite datasets of Table~\ref{tab:datasets_all} the rejection fires on at most $3\%$ of draws (com-amazon, the most triangle-dense of the set), and never on a bipartite host, where $\Gamma(v_1)\cap\Gamma(v_2)=\emptyset$ for every edge.
    \end{proof}

    It remains to implement the weighted edge draw of step~\emph{(i)} efficiently. A textbook solution stores the prefix sums of the $w_e/N(P_4)$ and binary-searches a uniform random value for each draw, costing $O(\log m)$ per sample for a graph with $m$ edges. Because BiLight draws up to millions of paths, we instead use
    \textit{Walker's Alias Method}~\cite{walker1977alias}, as described in Algorithm~\ref{alg:preprocess}, which reduces the cost of each draw to $O(1)$.
    Let $p:E\to [0,1]$ be a probability distribution. To draw an edge from $E=\{e_1,\dots, e_m\}$ we pick $i\in\{1,\ldots,m\}$ and a value $r\in[0,1)$, both uniformly at random. If $r < \mathit{prob}[i]$ we return the edge $e_i$, otherwise we return its alias, the edge $e_{\mathit{alias}[i]}$.
    A single index draw plus one comparison thus return an edge with the intended probability $p(e)$.

\section{Sampling Algorithm}\label{sec:samplingAlgo}

    \begin{table*}[htbp]
        \centering
        \scriptsize
        \setlength{\tabcolsep}{4pt}
        \renewcommand{\arraystretch}{1.35}
        \begin{tabular}{@{}lp{\dimexpr\linewidth-4.5em}@{}}
            \toprule
            \textbf{Motif} & \textbf{$\operatorname{ext}(M,\,P)$} \\ \midrule
            M1 & $\mathbb{I}_{c_4} \cdot \sum_{u \in I_L} |\Gamma(u) \cap I_R|$ \\
            M2 & $\mathbb{I}_{c_4} \cdot \left( \binom{|I_L|}{2} + \binom{|I_R|}{2} \right)$ \\
            M3 & $E(I_L, I_R) + \mathbb{I}_{c_4} \cdot \left( \sum_{u \in I_L} \left( |\Gamma(u) \cap \Gamma'(v_0)| + |\Gamma(u) \cap \Gamma'(v_2)| \right) + \sum_{u \in I_R} \left( |\Gamma(u) \cap \Gamma'(v_1)| + |\Gamma(u) \cap \Gamma'(v_3)| \right) + |I_L||I_R| - |I_L \cap I_R| \right)$ \\
            M4 & $\binom{|I_L|}{2} + \binom{|I_R|}{2} + \mathbb{I}_{c_4} \cdot \left( |I_L|(d'(v_1)+d'(v_3)-2) + |I_R|(d'(v_0)+d'(v_2)-2) \right)$ \\
            M5 & $\sum_{u \in I_L} |\Gamma'(u) \cap \Gamma'(v_2)| + \sum_{u \in I_R} |\Gamma'(u) \cap \Gamma'(v_1)| + \mathbb{I}_{c_4} \cdot \Bigl( \sum_{u \in I_L} d'(u) + \sum_{u \in I_R} d'(u)$ \newline ${}+ \sum_{u \in I_L} \left( d'(v_0)+d'(v_2) - \mathbb{I}(u\in\Gamma(v_0)) - \mathbb{I}(u\in\Gamma(v_2)) \right) + \sum_{u \in I_R} \left( d'(v_1)+d'(v_3) - \mathbb{I}(u\in\Gamma(v_1)) - \mathbb{I}(u\in\Gamma(v_3)) \right) \Bigr)$ \\
            M6 & $E(I_R, \Gamma'(v_3)) + E(I_L, \Gamma'(v_0)) + |I_L||I_R| - |I_L \cap I_R| + \mathbb{I}_{c_4} \cdot \left( E(\Gamma'(v_1), \Gamma'(v_0)) + E(\Gamma'(v_1), \Gamma'(v_2)) + E(\Gamma'(v_3), \Gamma'(v_0)) + E(\Gamma'(v_3), \Gamma'(v_2)) \right)$ \\
            M7 & $|I_L|(d'(v_1)-1) + |I_R|(d'(v_2)-1) + \mathbb{I}_{c_4} \cdot \sum_{i=0}^{3} \binom{d'(v_i)}{2}$ \\
            M8 & $\sum_{u \in \Gamma'(v_1)} |\Gamma(u) \cap \Gamma'(v_2)| + \sum_{u \in I_L} d'(u) + \sum_{u \in I_R} d'(u) + \sum_{u \in I_L}\!\left( d'(v_2) - \mathbb{I}(u\in\Gamma(v_2)) \right) + \sum_{u \in I_R}\!\left( d'(v_1) - \mathbb{I}(u\in\Gamma(v_1)) \right)$ \newline ${}+ \mathbb{I}_{c_4} \cdot \bigl( d'(v_0)d'(v_1) + d'(v_1)d'(v_2) + d'(v_2)d'(v_3) + d'(v_3)d'(v_0) - |\Gamma'(v_0)\cap\Gamma'(v_1)| - |\Gamma'(v_1)\cap\Gamma'(v_2)| - |\Gamma'(v_2)\cap\Gamma'(v_3)| - |\Gamma'(v_3)\cap\Gamma'(v_0)| \bigr)$ \\
            M9 & $|I_L|(d'(v_3)-1) + |I_R|(d'(v_0)-1) + \mathbb{I}_{c_4} \cdot \left( d'(v_1)d'(v_3) - |I_L| + d'(v_0)d'(v_2) - |I_R| \right)$ \\
            M10 & $\sum_{u \in I_L}\!\left( d'(v_0) - \mathbb{I}(u\in\Gamma(v_0)) \right) + \sum_{u \in I_R}\!\left( d'(v_3) - \mathbb{I}(u\in\Gamma(v_3)) \right) + E(\Gamma'(v_0), \Gamma'(v_1)) + E(\Gamma'(v_3), \Gamma'(v_2)) + \mathbb{I}_{c_4} \cdot \sum_{i=0}^{3} \sum_{u \in \Gamma'(v_i)} d'(u)$ \\
            M11 & $\sum_{u \in \Gamma'(v_0)} |\Gamma(u) \cap \Gamma'(v_3)|$ \\
            M13 & $\binom{d'(v_1)}{2} + \binom{d'(v_2)}{2}$ \\
            M14 & $d'(v_1)d'(v_2) - |\Gamma'(v_1) \cap \Gamma'(v_2)|$ \\
            M15 & $\sum_{u \in \Gamma'(v_1)} d'(u) + \sum_{u \in \Gamma'(v_2)} d'(u) + \left( d'(v_1)d'(v_0) - |\Gamma'(v_1)\cap\Gamma'(v_0)| \right) + \left( d'(v_2)d'(v_3) - |\Gamma'(v_2)\cap\Gamma'(v_3)| \right)$ \\
            M16 & $\binom{d'(v_0)}{2} + \binom{d'(v_3)}{2} + d'(v_1)d'(v_3) - |I_L| + d'(v_0)d'(v_2) - |I_R|$ \\
            M17 & $\left( d'(v_0)d'(v_3) - |\Gamma'(v_0)\cap\Gamma'(v_3)| \right) + \sum_{u \in \Gamma'(v_0)} d'(u) + \sum_{u \in \Gamma'(v_3)} d'(u)$ \\ \bottomrule
        \end{tabular}
        \caption{Local counting patterns for a \emph{general} host (bipartite or non-bipartite): the closed-form expression for $\operatorname{ext}(M,P)$ for each motif.}
        \label{tab:counting_logic_gen}
    \end{table*}
In Algorithm~\ref{alg:estimate}, $\operatorname{ext}(M,P)$ denotes the number of ways a sampled $P_4$ can be extended to an occurrence of motif~$M$ in~$G$, and $c_M$ is the number of distinct copies of $P_4$ inside a single instance of~$M$.
    Many of the expressions in Table~\ref{tab:counting_logic_gen} have two additive components: a \emph{path term} that is always evaluated, and a \emph{cycle term} multiplied by~$\mathbb{I}_{c_4}$, which contributes only when the seed path closes into a~$C_4$.
    Motif~$M_{12}$ is the six-vertex star $K_{1,5}$; its count is obtained directly from the degree sequence as $\sum_{v}\binom{d(v)}{5}$, so it is excluded from sampling.

        We relabel the vertices of the $P_4$ sampled in Algorithm~\ref{alg:estimate} in path order as $P = (v_0, v_1, v_2, v_3)$, with $v_0$ and $v_3$ the endpoints, and compute $\operatorname{ext}(M, P)$ for each motif~$M$.
    Each $P_4$ can potentially be extended to a 6-vertex motif by adding two more vertices from the neighborhoods of the path vertices.
    We derive a closed-form expression for each motif; Table~\ref{tab:counting_logic_gen} lists all of them.
    These expressions use neighborhood intersections, edge counts between vertex sets, and vertex degrees.

    We use the following auxiliary definitions in Table \ref{tab:counting_logic_gen}.
    For a vertex~$u$, let $\Gamma'(u) = \Gamma(u) \setminus \{v_0,v_1,v_2,v_3\}$ be the neighbors of~$u$ that lie outside the path, and $d'(u) = |\Gamma'(u)|$.
    We define two \emph{filtered intersection sets}:
    \begin{align*}
        I_L &= \Gamma(v_1) \cap \Gamma(v_3) \setminus \{v_0, v_2\}, \\
        I_R &= \Gamma(v_0) \cap \Gamma(v_2) \setminus \{v_1, v_3\}.
    \end{align*}
    Let the indicator $\mathbb{I}_{c_4}$ equals~$1$ when $v_0$ is adjacent to~$v_3$, and $0$ otherwise. Similarly, $\mathbb{I}(Q)$ represents the indicator variable for a proposition $Q.$
    We also use the notation $E(U,W) := |\{(u,w)\in E : u\in U,\,w\in W\}|$.
    We compute this efficiently by iterating over the \emph{smaller} set and probing the \emph{larger} set's adjacency, stored as a hash set.
    The same asymmetry-aware strategy applies to all intersection and edge-count computations throughout the local counting: when comparing two sets of different sizes, we always iterate over the smaller one.

    Although our motifs are bipartite, neither the sampler nor the counting patterns assume a bipartite host. The edge weight $w_e$ of Algorithm~\ref{alg:preprocess} and the rejection step in the proof of Claim~\ref{clm:uniformrandom} already discount the triangles through each central edge, so a uniformly random $P_4$ is drawn on any input graph. The patterns of Table~\ref{tab:counting_logic_gen} are stated in the same generality: the adjacency indicators $\mathbb{I}(u\in\Gamma(v_i))$ and the intersection corrections such as $-|I_L\cap I_R|$ and $-|\Gamma'(v_i)\cap\Gamma'(v_j)|$ exclude the same-side adjacencies and triangles that a bipartite host rules out by construction. On a bipartite host every one of these terms is zero, and each expression reduces to its simpler bipartite form.

    \begin{algorithm}[htbp]
    \DontPrintSemicolon
    \caption{BiLight: Estimation}\label{alg:estimate}
    \KwIn{$G$; Arrays $\mathit{prob}[i]$, $\mathit{alias}[i]$ for $i=1,\dots,m$ and $N(P_4)$ from Algorithm~\ref{alg:preprocess}; sample budget $k$}
    \KwOut{Estimated counts $\widehat{N}(M_1),\ldots,\widehat{N}(M_{17})$}
    $s_j \gets 0$ for $j\neq 12$;\enspace $\widehat{N}(M_{12}) \gets \sum_{v}\binom{d(v)}{5}$\;
    \For{$t \gets 1$ \KwTo $k$}{
      Sample edge $e=(u,v)$ with prob. $w_e/N(P_4)$\;
      \Repeat{$u' \neq v'$}{
        Pick $u' \in \Gamma(u)\setminus\{v\}$ uniformly at random\;
        Pick $v' \in \Gamma(v)\setminus\{u\}$ uniformly at random\;
      }\tcp*{$u'=v'$ closes a triangle, not a $P_4$}
      $P \gets (u',\,u,\,v,\,v')$\;
      \ForEach{motif $M_j$, $j\neq 12$}{
        $s_j \gets s_j + \operatorname{ext}(M_j,\,P)\;/\;c_{M_j}$\;
      } \tcp*{Table~\ref{tab:counting_logic_gen} provides $\operatorname{ext}(M_j,\,P)$ }
    }
    \For{$j \gets 1$ \KwTo $17$, $j\neq 12$}{
      $\widehat{N}(M_j) \gets \dfrac{N(P_4)}{k}\cdot s_j$\;
    }
    \Return{$\widehat{N}(M_1),\ldots,\widehat{N}(M_{17})$}\;
    \end{algorithm}

\begin{table*}[htbp]
        \centering
        \footnotesize
        \begin{tabular}{@{}lrrrrc@{}}
            \toprule
            \textbf{Dataset} & \textbf{Vertices} & \textbf{Edges} & \textbf{$P_4$ Count} & \textbf{Runtime ($k{=}5$K)} & \textbf{Source} \\ \midrule
            \multicolumn{6}{@{}l}{\textit{Bipartite}} \\ \midrule
            Amazon (Electronics) & 4.5M & 7.7M & 25.8B & 12\,s & \cite{datasource-amazon, mcauley2015image, he2016ups} \\
            Amazon (Books) & 10.0M & 22.3M & 352.0B & 72\,s & \cite{datasource-amazon, mcauley2015image, he2016ups} \\
            IMDB & 1.2M & 3.8M & 4.6B & 2\,s & \cite{datasource-imdb} \\
            Flickr & 470.3K & 8.5M & 2.6T & 232\,s & \cite{datasource-flickr} \\ \midrule
            \multicolumn{6}{@{}l}{\textit{Additional non-bipartite}} \\ \midrule
            wordassociation & 10.6K & 63.8K & 79.2M & 1\,s & \cite{datasource-law, boldi2004webgraph, boldi2011layered} \\
            facebook-wall & 44.0K & 182.4K & 126.3M & 1\,s & \cite{datasource-mpisws, viswanath2009evolution} \\
            facebook-links & 63.4K & 816.9K & 7.5B & 9\,s & \cite{datasource-mpisws, viswanath2009evolution} \\
            com-amazon.ungraph & 334.9K & 925.9K & 81.0M & $<$1\,s & \cite{datasource-snap} \\
            com-dblp.ungraph & 317.1K & 1.0M & 675.6M & 4\,s & \cite{datasource-snap} \\
            web-BerkStan      & 654.8K & 6.6M & 1.7T & 340\,s & \cite{datasource-snap} \\
            as-skitter & 1.7M & 11.1M & 1.4T & 350\,s & \cite{datasource-snap} \\
            cit-Patents & 3.8M & 16.5M & 9.1B & 3\,s & \cite{datasource-snap} \\
            livejournal & 5.4M & 49.5M & 2.6T & 106\,s & \cite{datasource-law, boldi2004webgraph, boldi2011layered} \\
            hollywood & 1.1M & 56.3M & 60.2T & 3,326\,s & \cite{datasource-law, boldi2004webgraph, boldi2011layered} \\
            soc-orkut & 3.0M & 106.3M & 16.3T & 298\,s & \cite{datasource-networkrepo} \\
            orkut-links & 3.1M & 117.2M & 22.2T & 564\,s & \cite{datasource-orkutlinks, mislove2007measurement} \\ \bottomrule
        \end{tabular}
        \caption{Datasets used in experiments. Vertex and edge counts refer to the largest connected component (LCC) where applicable. $P_4$ Count is the number of $P_4$ subgraphs in the graph (sampling space). Runtime ($k{=}5$K) is BiLight's estimation time at sample budget $k=5$K in seconds, excluding the one-time graph setup; bipartite values are averages over independent runs. The Flickr file is a mixed user--user and user--group graph; we keep only the user--group edges, which makes it bipartite.}
        \label{tab:datasets_all}
    \end{table*}

    \begin{lemma}\label{lem:unbiased}
    For any motif $M=M_i$ with $i\neq 12$, the estimator is unbiased, $\mathbb{E}[\widehat{N}(M)]=N(M)$, and
 \[
        \mathbb{V}ar[\widehat{N}(M)]\le \frac{1}{k}\,\frac{N(P_4)}{c_M}\bigl(N(M)+p\bigr),
    \]
    where $p$ is the number of ordered pairs of $M$-instances that share a $P_4$ seed. Thus, our estimator yields an $(\epsilon, \delta)$-estimator by
taking the median of $O(\log(1/\delta))$ estimators.
    \end{lemma}
    \begin{proof}
    Let $\mathcal{P}$ denote the set of all $P_4$ subgraphs in~$G$, so $|\mathcal{P}| = N(P_4)$.
    The estimator for the non-induced count of motif $M$ scales the average local count by the total number $N(P_4)$ of paths:
    \[
        \widehat{N}(M) = \frac{N(P_4)}{k}\sum_{t=1}^{k}\frac{\operatorname{ext}(M,P_t)}{c_M}
    \]
    By Claim~\ref{clm:uniformrandom}:
    \begin{align*}
        \mathbb{E}[\widehat{N}(M)]
          &= \frac{N(P_4)}{k}\sum_{t=1}^{k}\mathbb{E}\!\left[\frac{\operatorname{ext}(M,P_t)}{c_M}\right] \\
          &= \frac{N(P_4)}{c_M}\sum_{P\in\mathcal{P}}\Pr(P)\cdot\operatorname{ext}(M,P) \\
          &= \frac{1}{c_M}\sum_{P\in\mathcal{P}}\operatorname{ext}(M,P)=N(M).
    \end{align*}
    The last equality follows from the fact that
    every instance of motif~$M$ in~$G$ contains exactly $c_M$ distinct copies of~$P_4$, 
    thus yielding $\sum_{P\in\mathcal{P}}\operatorname{ext}(M,P) = c_M \cdot N(M)$.

    For the variance, consider a single draw $Y=N(P_4)\operatorname{ext}(M,P)/c_M$; as the $k$ paths are drawn independently (with replacement), $\widehat{N}(M)$ is the average of $k$ i.i.d.\ copies of~$Y$. Let $X_i$ indicate that the $i$'th copy of $M$, $1\leq i\leq N(M)$, contains the drawn path.
    \begin{align}
        \mathbb{V}ar[Y]
          &= \mathbb{V}ar\Bigl[\frac{N(P_4)}{c_M}\sum_{i=1}^{N(M)} X_i\Bigr] \notag\\
          &= \frac{N(P_4)^2}{c_M^2}\,\mathbb{V}ar\Bigl[\sum_{i=1}^{N(M)} X_i\Bigr] \notag\\
          &= \frac{N(P_4)^2}{c_M^2}\Bigl[\sum_{i=1}^{N(M)} \mathbb{V}ar[X_i] 
          + \sum_{i\neq j}\mathbb{C}ov(X_i,X_j)\Bigr] \notag\\
          &= \frac{N(P_4)^2}{c_M^2}\Bigl[N(M)\Bigl(\frac{c_M}{N(P_4)}-\frac{c_M^2}{N(P_4)^2}\Bigr) \notag\\
          &\quad + \sum_{i\neq j}\bigl(\mathbb{E}[X_iX_j]-\mathbb{E}[X_i]\,\mathbb{E}[X_j]\bigr)\Bigr].
        \label{eq:var_cov}
    \end{align}
    We only consider the pairs $\{X_i,X_j\}$ that share at least one seed path, since otherwise $\mathbb{E}[X_iX_j]=0$ and $\mathbb{C}ov(X_i,X_j)<0$. If instances $i,j$ share $a_{ij}$ seed paths, then
    $\mathbb{E}[X_iX_j]=\Pr[X_i{=}1]\Pr[X_j{=}1|X_i{=}1]=\frac{c_M}{N(P_4)}\frac{a_{ij}}{c_M}=\frac{a_{ij}}{N(P_4)}$, so
    \[
        \mathbb{C}ov(X_i,X_j)=\frac{a_{ij}}{N(P_4)}-\frac{c_M^2}{N(P_4)^2}.
    \]
    Using $a_{ij}\le c_M$ and ignoring the negative contributions, we obtain
    \[
        \mathbb{V}ar[Y]\le \frac{N(P_4)}{c_M}\bigl(N(M)+p\bigr),
    \]
    and $\mathbb{V}ar[\widehat{N}(M)]=\mathbb{V}ar[Y]/k$ by independence, which is the stated bound.
  Let $k=\frac{8N(P_4)(N(M)+p)}{c_M\epsilon^2N(M)^2}$. Chebyshev's inequality then yields
  \begin{align*}
      \Pr[|\widehat{N}(M)-N(M)|\geq \epsilon N(M)]\leq \frac{\mathbb{V}ar[\widehat{N}(M)]}{\epsilon^2N(M)^2}\\
      \leq\frac{N(P_4)(N(M)+p)}{c_Mk\epsilon^2N(M)^2}=\frac{1}{8}.
  \end{align*}
So, we can turn this estimator into an $(\epsilon, \delta)$-estimator by
taking the median of $O(\log(1/\delta))$ estimators.

  \end{proof}

\begin{table*}[htbp]
        \centering
        \scriptsize
        \setlength{\tabcolsep}{1.8pt}
        \caption{Motif induced counts on the bipartite datasets: BiLight ($k=100$K) vs.\ Motivo (50M color-coding samples).
 Missing entries indicate that Motivo did not report the motif at that budget.}
        \label{tab:motif_counts}
        \begin{tabular}{@{}llccccccccccccccccr@{}}
            \toprule
            \textbf{Dataset} & \textbf{Method} & M1 & M2 & M3 & M4 & M5 & M6 & M7 & M8 & M9 & M10 & M11 & M13 & M14 & M15 & M16 & M17 & \textbf{Runtime} \\
            \midrule
            \multirow{2}{*}{\shortstack[l]{Amazon\\Elec.}} & BiLight & 3.62e6 & 1.07e12 & 2.45e8 & 5.36e13 & 3.76e10 & 2.98e9 & 8.70e14 & 3.57e12 & 5.92e14 & 8.42e11 & 1.24e10 & 2.22e17 & 1.21e15 & 5.94e14 & 6.01e16 & 1.04e14 & 4\,min \\
            & Motivo & --- & --- & --- & 2.72e12 & --- & --- & 3.72e14 & 1.54e12 & 1.55e14 & 2.21e11 & --- & 2.17e17 & 1.24e15 & 6.09e14 & 5.35e16 & 9.97e13 & 5\,h 3\,min \\
            \midrule
            \multirow{2}{*}{\shortstack[l]{Amazon\\Books}} & BiLight & 3.52e9 & 1.25e13 & 1.10e11 & 5.45e14 & 1.21e13 & 8.78e11 & 1.96e16 & 7.93e14 & 5.10e15 & 1.36e14 & 1.32e12 & 2.12e19 & 6.72e17 & 8.77e16 & 4.34e17 & 5.23e15 & 23\,min \\
             & Motivo & --- & 1.15e13 & --- & 4.66e14 & 1.78e13 & --- & 1.96e16 & 9.71e14 & 4.97e15 & 9.17e13 & --- & 2.41e19 & 8.40e17 & 1.17e17 & 5.09e17 & 7.12e15 & 12\,h 57\,min \\
            \midrule
            \multirow{2}{*}{IMDB} & BiLight & 1.06e8 & 5.84e9 & 1.13e9 & 1.27e11 & 1.93e10 & 5.44e9 & 1.56e12 & 4.58e11 & 9.68e11 & 2.84e11 & 1.34e10 & 7.60e13 & 1.24e13 & 1.95e13 & 4.14e13 & 9.73e12 & 45\,s \\
            & Motivo & 1.16e8 & 5.07e9 & 1.14e9 & 1.13e11 & 1.86e10 & 5.34e9 & 1.51e12 & 4.51e11 & 9.22e11 & 2.79e11 & 1.30e10 & 7.50e13 & 1.22e13 & 1.94e13 & 4.05e13 & 9.67e12 & 16\,min \\
            \midrule
            \multirow{2}{*}{Flickr} & BiLight & 2.46e13 & 1.55e16 & 5.87e14 & 3.34e17 & 1.22e16 & 2.96e15 & 3.26e18 & 1.85e17 & 1.16e18 & 6.21e16 & 2.32e15 & 7.17e19 & 1.45e18 & 1.66e18 & 1.36e19 & 4.41e17 & 1\,h 17\,min \\
            & Motivo & 1.98e13 & 1.69e16 & 5.86e14 & 3.47e17 & 1.22e16 & 3.00e15 & 3.30e18 & 1.87e17 & 1.18e18 & 6.26e16 & 2.31e15 & 7.16e19 & 1.46e18 & 1.65e18 & 1.36e19 & 4.40e17 & 5\,h 34\,min \\
            \bottomrule
        \end{tabular}
    \end{table*}
  \begin{claim}\label{clm:runtime}
   For a graph with maximum degree $\Delta$, arboricity $\alpha$, $n$ vertices and $m$ edges,
   the running time and total storage (including all preprocessing) of BiLight are $O(\alpha m + k\Delta^2)$ and $O(n+m)$,
respectively.
  \end{claim}
\begin{proof}
    Preprocessing (Algorithm~\ref{alg:preprocess}) builds the edge weights and the alias table. The weight of each edge requires $|\Gamma(u)\cap\Gamma(v)|$, so preprocessing costs $O\!\left(\sum_{(u,v)\in E}\min(d(u),d(v))\right)$, the usual bound for enumerating triangles, which is $O(\alpha m)$ for a graph of arboricity~$\alpha$~\cite{chiba1985arboricity}. Our implementation first tests bipartiteness in $O(n+m)$ and skips the intersections on a bipartite host, where they are all empty, reducing this step to $O(m)$. This is a one-time cost, independent of the sample budget~$k$. Each sample costs $O(1)$ to draw and $O(\Delta^2)$ to score: the patterns of Table~\ref{tab:counting_logic_gen} reduce to adjacency-list intersections, each $O(\Delta)$ with hash sets, and those that sum an intersection over a neighbor set (e.g., M11) need $O(\Delta)$ of them. The total time for $k$ samples is therefore $O(\alpha m + k\Delta^2)$, \emph{linear} in the sample budget, whereas exact enumeration over 6-vertex tuples scales with $O(\Delta^4)$ or worse and is prohibitive on large graphs. Figure~\ref{fig:runtime} in Appendix~\ref{app:runtime} confirms the linear dependence on~$k$ empirically.

    The storage budget is independent of $k$. The adjacency-list representation of $G$ requires $O(n+m)$ space. Our implementation additionally keeps an explicit list of the $m$ edges when constructing the edge weight $w_e$ for each edge $e$. The two alias arrays $\mathit{prob}[\cdot]$ and $\mathit{alias}[\cdot]$ add $O(m)$ words.
    Between samples, BiLight updates only a fixed set of motif accumulators (one per motif), and the temporary structures used during local counting (e.g., neighbor-set intersections) are $O(\Delta)$ in the worst case and are reused across the $k$ iterations rather than accumulated.
\end{proof}

\begin{remark}\label{rem:runtime}
    At small sample budgets, the running time is dominated by the one-time preprocessing.
\end{remark}
On IMDB, for instance, setup takes about 10\,s while estimation at $k=10$ takes under 30\,ms. Figure~\ref{fig:runtime} (Appendix~\ref{app:runtime}) plots estimation time alone, which grows linearly in~$k$ from the smallest budget; its small irregularities reflect that the cost of scoring a sampled $P_4$ varies with the degrees of its vertices, not a deviation from the $O(\alpha m+k\Delta^2)$ bound.

    \section{Experimental Evaluation}\label{sec:results}

    All experiments (BiLight and Motivo~\cite{bressan2019motivoarxiv, bressan2019motivo, bressan2021faster}) were run on a single machine: an Intel Core i7-10750H CPU at 2.60\,GHz (6 cores), 32\,GB RAM, running Windows~11 Enterprise.
    The BiLight implementation used to produce these results is available as
open source code at \href{https://github.com/celikmustaa/BiLight}{\nolinkurl{github.com/celikmustaa/BiLight}}.

    BiLight runs on any undirected graph and estimates the counts of the 17 \emph{bipartite} 6-vertex motifs as subgraphs of that graph. As listed in Table~\ref{tab:datasets_all}, we use four large bipartite datasets: Amazon Ratings (Books and Electronics), IMDB (actor--movie), and Flickr (user--group) and middle to large-scale nonbipartite graphs.  Table~\ref{tab:datasets_all} lists all datasets; the top block is bipartite, restricted to the largest connected component (LCC).

    \subsection{Induced and non-induced Motifs}

Using BiLight, we are able to obtain the counts of non-induced copies of
the motifs in Figure~\ref{fig:motifs} on bipartite as well as nonbipartite networks. As observed in~\cite{pinar2017escape}, one can formalize a linear relationship between induced and non-induced counts of motifs in a network.
We generalize this idea for our framework, by letting $\mathbf{C}$ and $\mathbf{N}$ denote the vectors with 17 entries,
where for each $i$, $C_i$ and $N_i$ denotes the number of induced and non-induced copies of motif $M_i$, respectively. We provide in Appendix~\ref{app:conversion_matrices}, the $17\times 17$ matrix $A$ such that $\mathbf{N} = A \mathbf{C}$ on any given bipartite network, where $A_{i,j}$ indicates the number of (non-induced) copies of motif $M_i$ appearing in an instance of motif $M_j$. Reversely, one can obtain $C$ given $N$ by the relation $\mathbf{C} = A^{-1} \mathbf{N}$. For a suitable motif ordering $A$ is lower triangular.
Using this conversion, we are able to obtain the counts of induced copies of each motif in Figure~\ref{fig:motifs} on bipartite networks, which we use for comparison of our results with state of the art algorithms.

        \begin{table}[!ht]
        \centering
        \footnotesize
        \caption{Average Distance Error on non-bipartite datasets (lower is better). All entries are proportions (e.g., $0.10 = 10\%$). Reference is BiLight's own estimate at $k=100$K.}
        \label{tab:accuracy}
        \begin{tabular}{@{}lccc@{}}
            \toprule
            \textbf{Dataset} & \textbf{$k=1$K} & \textbf{$k=5$K} & \textbf{$k=20$K} \\
            \midrule
            wordassociation     & $0.16$ & $0.05$ & $0.01$ \\
            facebook-wall       & $0.15$ & $0.05$ & $0.05$ \\
            facebook-links      & $0.10$ & $0.03$ & $0.02$ \\
            com-amazon          & $0.06$ & $0.03$ & $0.01$ \\
            com-dblp            & $0.06$ & $0.03$ & $0.00$ \\
            web-BerkStan        & $0.10$ & $0.08$ & $0.02$ \\
            as-skitter          & $0.05$ & $0.03$ & $0.02$ \\
            cit-Patents         & $0.04$ & $0.10$ & $0.02$ \\
            livejournal         & $0.07$ & $0.05$ & $0.03$ \\
            hollywood           & $0.05$ & $0.00$ & $0.01$ \\
            soc-orkut           & $0.22$ & $0.22$ & $0.08$ \\
            orkut-links         & $0.11$ & $0.12$ & $0.06$ \\
            \bottomrule
        \end{tabular}
    \end{table}

    \subsection{Performance and Comparison with previous work}\label{sec:motivo}

\begin{figure}[!t]
        \centering
        \includegraphics[width=0.95\linewidth]{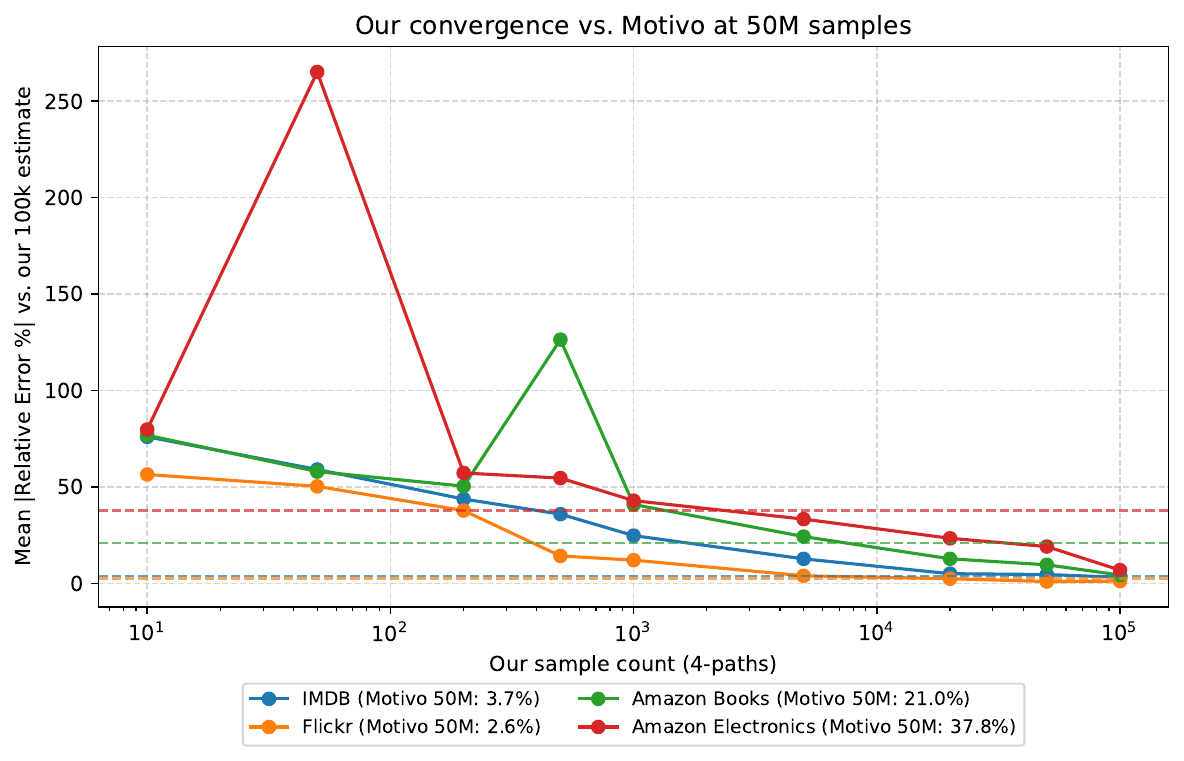}
        \caption{BiLight's convergence: the mean relative difference (\%) between BiLight's induced counts at budget $k$ and at $k=100$K, averaged over the 16 sampled motifs. Horizontal dashed lines: the mean relative difference between BiLight at $k=100$K and Motivo at 50M samples, averaged over the motifs Motivo reports (where applicable).}
        \label{fig:motivo_ours_samples}
    \end{figure}

    Since exact counting is infeasible, we define \emph{ground truth} as BiLight's estimate at $k=100$K. For the bipartite datasets, which have five independent runs at each budget, we average the runs' estimates at both the target and the reference budget before computing the error; for the non-bipartite single-run experiments, the $k=100$K value serves directly as the reference. We measure accuracy using the \textit{Average Distance Error}: the mean, over the 16 non-trivial motifs (M12 is excluded), of the per-motif relative error $|\widehat{N}(M_i)-N^{\mathrm{ref}}(M_i)|/N^{\mathrm{ref}}(M_i)$.
    Table~\ref{tab:accuracy} reports this error on the non-bipartite datasets at $k=1$K, $5$K, and $20$K. Nine of the twelve networks converge to within 8\% error at $k=5$K, and nine reach 4\% or better at $k=20$K; the two Orkut graphs, the largest in the set by edge count, remain the slowest to converge. On the bipartite datasets, the mean error at $k=50$K is below 10\% on all four graphs, with Flickr reaching 0.3\%. Repeatability tracks graph density: the run-to-run coefficient of variation of the individual motif estimates at $k=50$K averages $0.01$ on Flickr and $0.06$ on IMDB, rising to $0.12$ on Amazon Books and $0.18$ on Amazon Electronics, the sparsest of the four.

    Bressan et al.~\cite{bressan2018motif} addressed the problem of counting motifs beyond five vertices in general graphs with color coding, and later released Motivo~\cite{bressan2019motivoarxiv, bressan2019motivo, bressan2021faster}, which extends that approach with succinct data structures and adaptive sampling. Motivo estimates the frequencies of \emph{induced} motifs on 5 to 8 vertices in a single framework, in general (non-bipartite as well as bipartite) graphs.

    We ran both methods on the four bipartite datasets (IMDB, Amazon Ratings Books and Electronics, Flickr): BiLight with sample budgets from $k=1$K to $k=100$K, and Motivo~\cite{bressan2019motivoarxiv, bressan2019motivo, bressan2021faster} with 10K, 100K, 1M, 10M, and (where feasible) 50M color-coding samples.

\begin{figure}[H]
        \centering
        \includegraphics[width=0.88\linewidth]{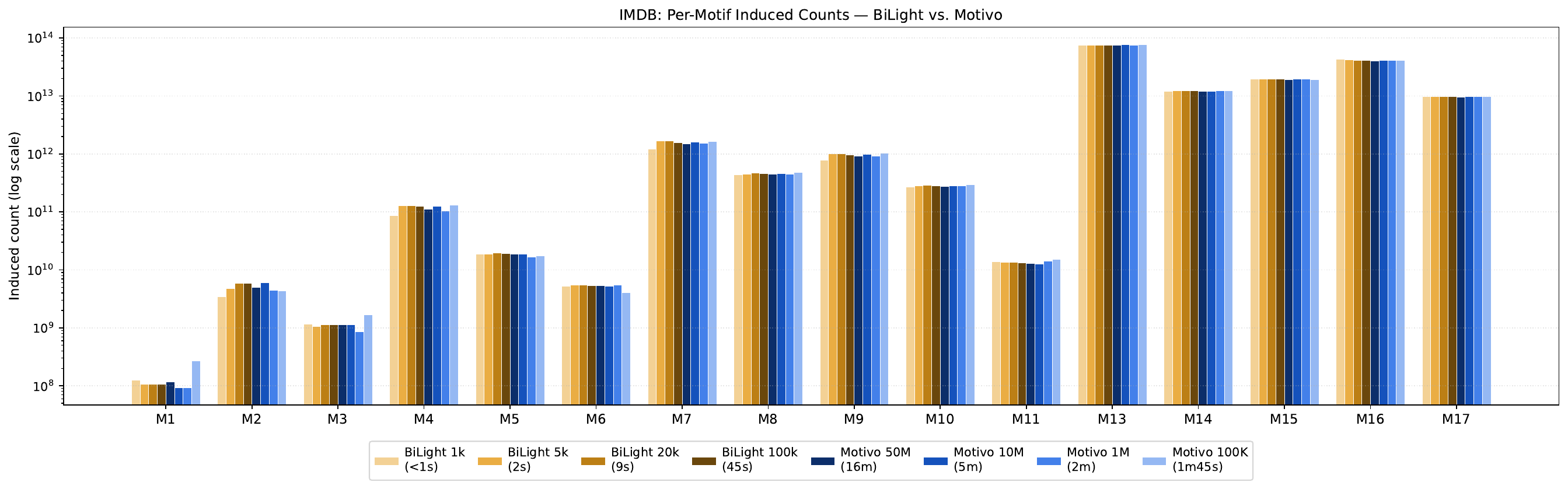}\\[0.05cm]
        \includegraphics[width=0.88\linewidth]{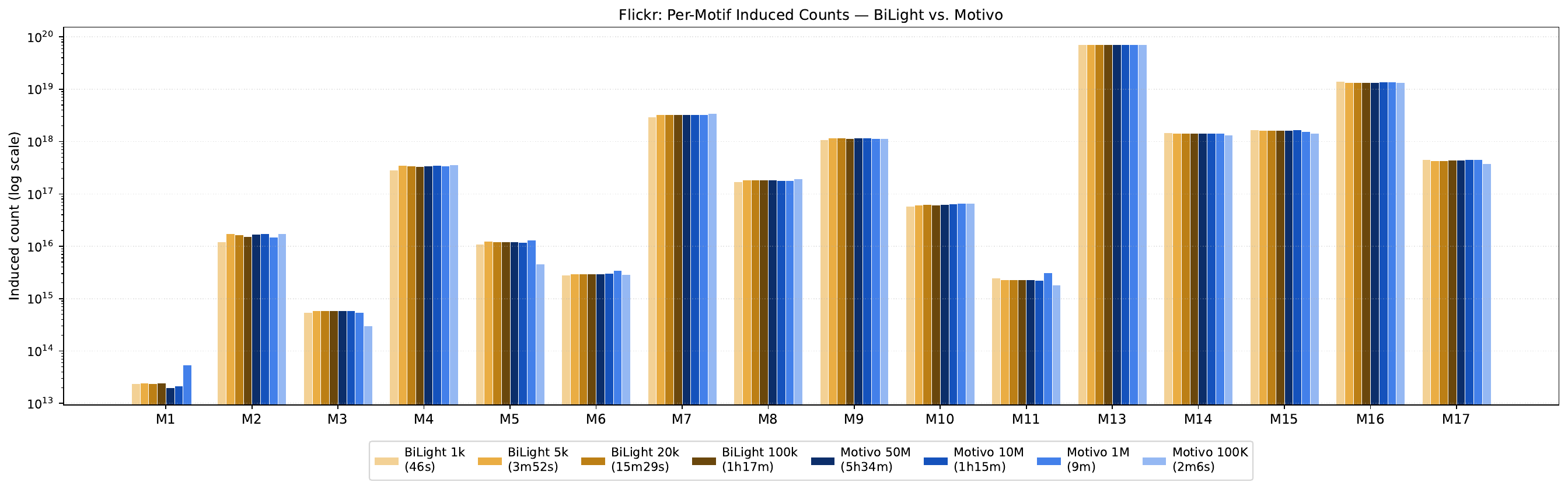}\\[0.05cm]
        \includegraphics[width=0.88\linewidth]{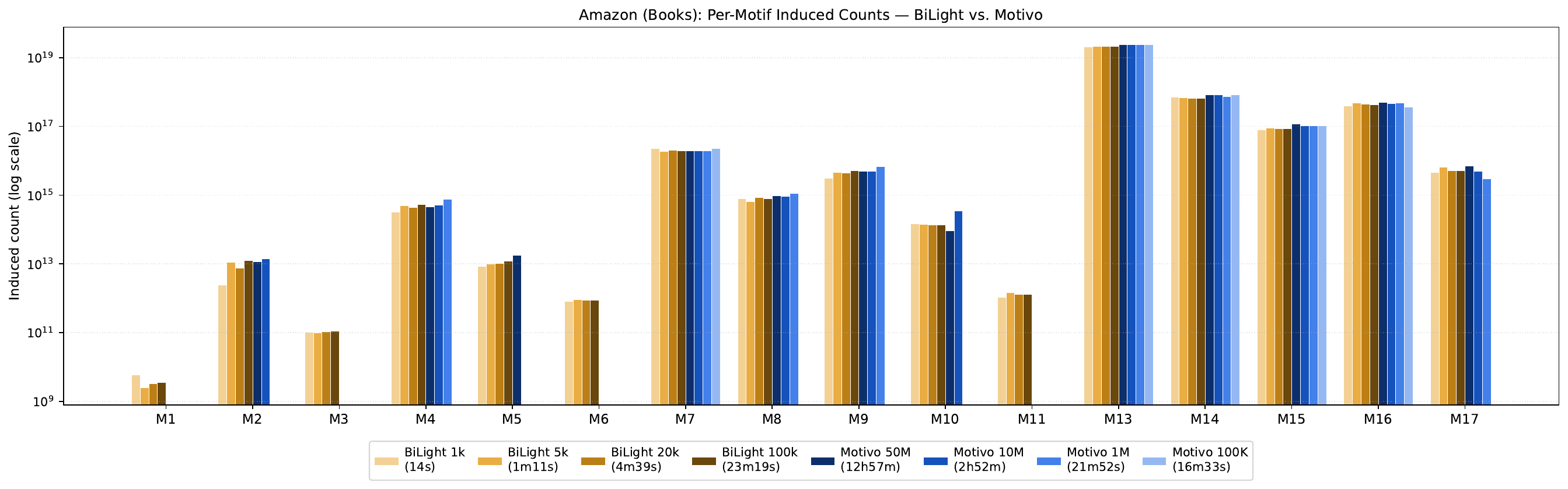}\\[0.05cm]
        \includegraphics[width=0.88\linewidth]{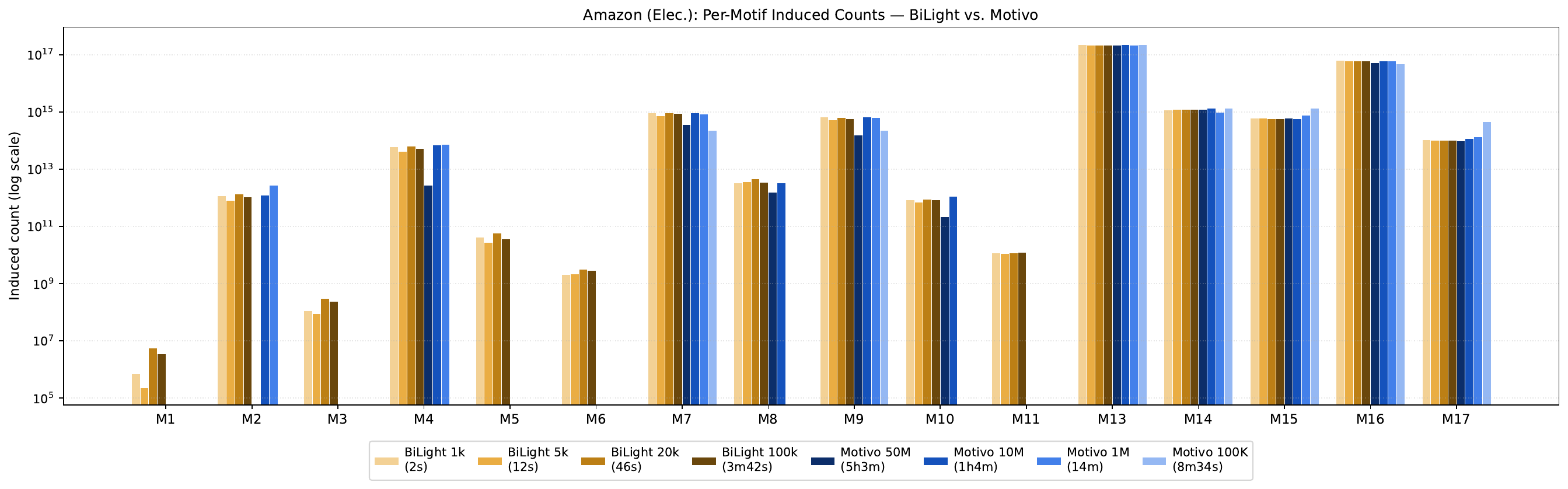}
        \caption{Per-motif induced counts at increasing sample budgets. \emph{Panels (top to bottom):} IMDB; Flickr; Amazon Ratings (Books); Amazon Ratings (Electronics). Teal bars = BiLight (light to dark: $k=1$K, 5K, 20K, 100K); orange bars = Motivo (dark to light: 50M, 10M, 1M, 100K color-coding samples). Wall-clock runtimes are shown in parentheses. Missing bars indicate motifs Motivo did not report at that budget. BiLight bars are averaged over five independent runs. The rarest motif, $M_1$ ($K_{3,3}$), is found on Amazon (Elec.) in only one of the five runs at $k{=}1$K and three of five at $k{=}5$K, so its bars at those budgets rest on a handful of hits; from $k{=}20$K on it is found in every run.}
        \label{fig:permotif_all}
    \end{figure}

    Table~\ref{tab:motif_counts} breaks down the per-motif induced counts: BiLight at $k=100$K (averaged over independent runs and converted via $A^{-1}$) against Motivo at 50M samples. M12 is omitted because BiLight obtains it analytically rather than by sampling.
    On IMDB and Flickr, all 16 comparable motifs agree closely, with only two above 10\% on IMDB and one on Flickr; on the Amazon datasets, Motivo leaves several motifs effectively unobserved even at 50M samples (M1, M3, M6 and M11 on Books, and additionally M2 and M5 on Electronics; on Electronics it returns $74$ for M3, against BiLight's $2.45\times 10^{8}$), while BiLight reports all of them. Figure~\ref{fig:permotif_all} (Appendix~\ref{app:permotif}) traces the same comparison across budgets: on IMDB and Flickr the darkest BiLight and Motivo bars ``meet in the middle,'' visually confirming that the two methods converge to the \emph{same} induced counts as their budgets grow, while on the Amazon datasets the gap persists on the motifs Motivo barely hits.


    \begin{figure*}[!t]
        \centering
        \includegraphics[width=0.95\textwidth]{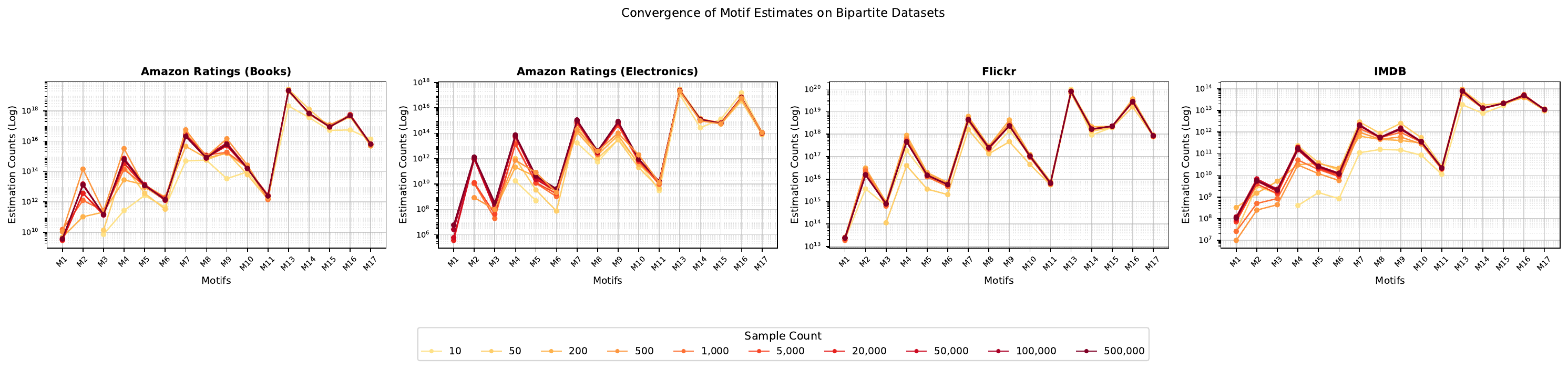}
        \caption{Convergence of motif count estimates on the four bipartite datasets (one subplot per dataset). $x$-axis = motif index (M12 excluded), $y$-axis = estimated count (log scale), line color = sample budget $k$ (lighter = smaller $k$).}
        \label{fig:convergence}
    \end{figure*}

    \begin{figure*}[!t]
        \centering
        \includegraphics[width=0.95\textwidth]{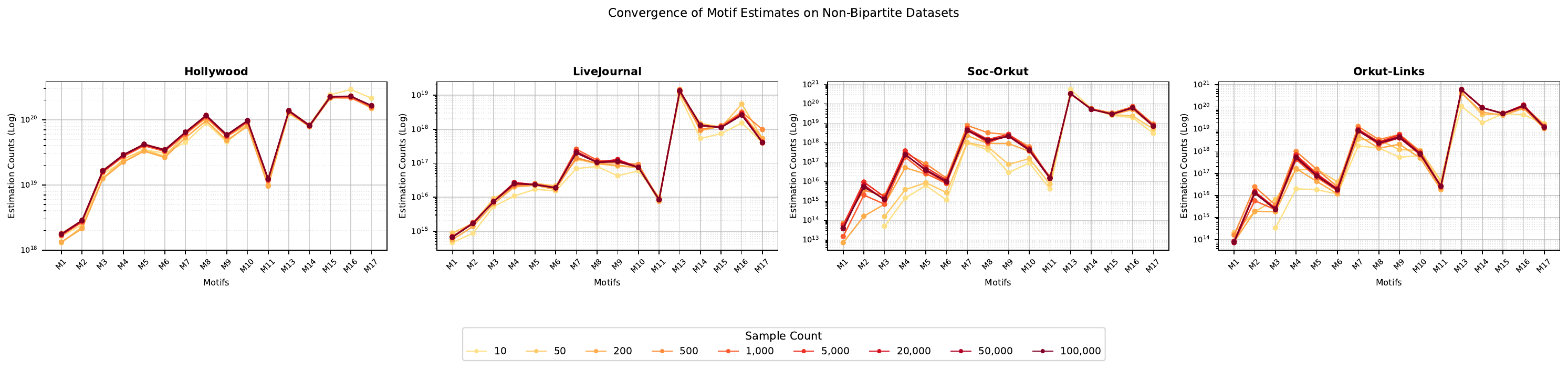}
        \caption{Convergence of motif count estimates on the four largest non-bipartite networks; layout as in Figure~\ref{fig:convergence}.}
        \label{fig:convergence_nonbipartite}
    \end{figure*}

\begin{figure}[!t]
        \centering
        \includegraphics[width=0.95\linewidth]{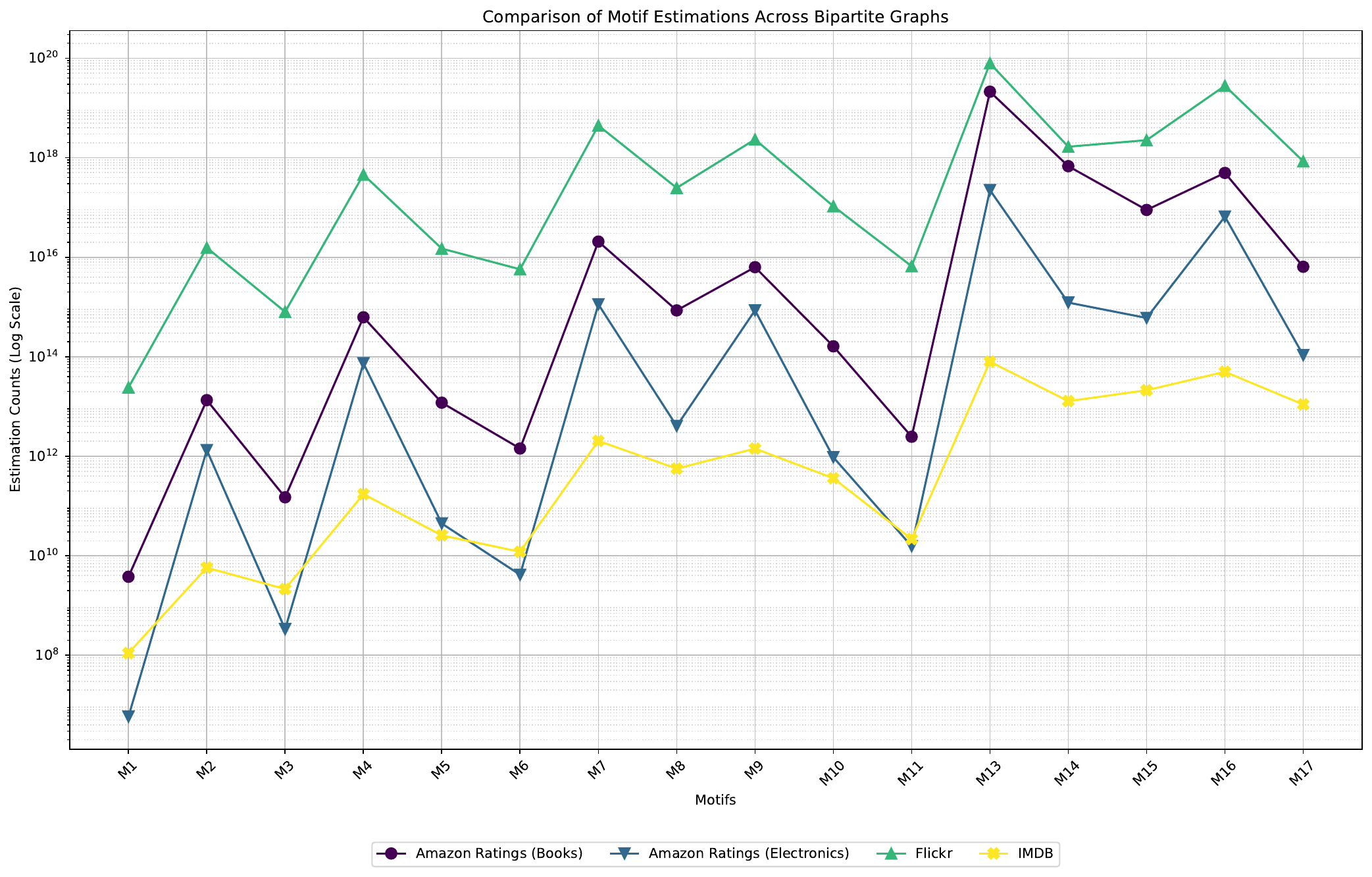}
        \caption{Estimated motif profiles (non-induced) of the four bipartite datasets at $k=500$K (one line per dataset).}
        \label{fig:profiles_bipartite}
    \end{figure}

    \begin{figure}[htbp]
        \centering
        \includegraphics[width=0.95\linewidth]{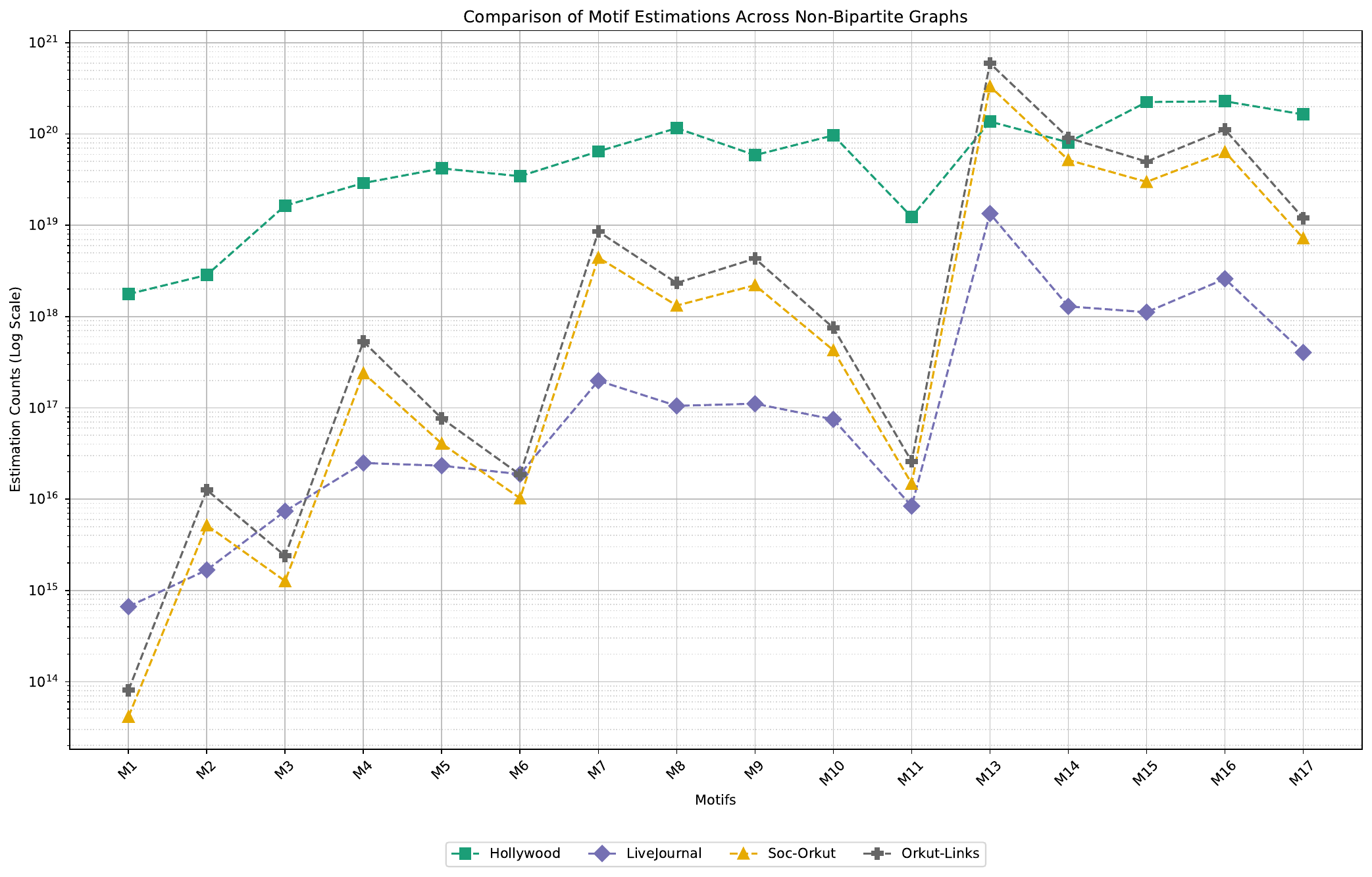}
        \caption{Estimated motif profiles (non-induced) of the four largest non-bipartite networks at $k=100$K.}
        \label{fig:profiles_nonbipartite}
    \end{figure}

    Figure~\ref{fig:motivo_ours_samples} aggregates these per-motif differences into a single curve per dataset. For a given budget~$k$ we convert BiLight's non-induced estimates to induced counts via $A^{-1}$ and plot the mean, over the 16 sampled motifs, of the relative difference between the estimate at~$k$ and the estimate at $k=100$K; each dashed line gives, for one dataset, the mean of $|C^{\text{BiLight}}_i - C^{\text{Motivo}}_i| / C^{\text{BiLight}}_i$ at $k=100$K over the motifs Motivo reports. The percentages quoted below are values of this second quantity. The curves fall as $k$ increases; on IMDB and Flickr, where Motivo reports all 16 non-trivial motifs, BiLight at $k=100$K agrees with their 50M-sample run to within 4\% ($3.7\%$ and $2.6\%$). On the Amazon datasets, their coverage is incomplete even at 50M samples (12/16 for Books, 10/16 for Electronics), and the agreement on the motifs they do report is uneven. On Books the mean difference is about $21\%$. On Electronics, the sparsest of the four, the two methods agree to within $4\%$ on the abundant tree-like motifs ($M_{13}$--$M_{15}$ and $M_{17}$) but differ by a factor of two or more on the rarer cycle-containing ones ($M_4$, $M_7$--$M_{10}$), where Motivo's 50M samples yield few hits; BiLight's estimates for those motifs stay stable between $k=50$K and $k=100$K. This reflects the difficulty of these sparse bipartite structures for general-purpose color-coding methods.

 \subsubsection{Convergence}

    We analyzed the convergence of BiLight by increasing the sample budget from $k=10$ to $k=500$K. Figure~\ref{fig:convergence} shows the estimated counts of all 16 sampled motifs on the four bipartite datasets: as $k$ grows, the curves overlap and the variance shrinks, indicating strong convergence. Figure~\ref{fig:convergence_nonbipartite} repeats the experiment on the four largest non-bipartite networks of Table~\ref{tab:datasets_all} by edge count. The same $P_4$ sampling and local counting patterns apply unchanged, and the estimates stabilize in the same way, confirming that BiLight is not limited to bipartite inputs.

    \subsubsection{Runtime efficiency:} The runtime column of Table~\ref{tab:motif_counts} summarizes the runtime gap: BiLight at $k=100$K finishes in $45$\,s on IMDB and in at most $1$\,h $17$\,min on Flickr, while Motivo~\cite{bressan2019motivoarxiv, bressan2019motivo, bressan2021faster} at 50M samples ranges from $16$\,min to nearly $13$\,h on the same inputs. To our knowledge, BiLight is the first framework that estimates all seventeen 6-vertex bipartite motifs efficiently on large networks.

    \subsubsection*{Trends in motif counts:}

    Figures~\ref{fig:profiles_bipartite} and~\ref{fig:profiles_nonbipartite} summarize the estimated motif profiles at the largest budgets ($k=500$K and $k=100$K, respectively). The profiles follow similar trends across datasets (up to scale), suggesting comparable relative motif importance despite the different domains; the non-bipartite networks sit several orders of magnitude higher owing to their larger edge densities and $P_4$ counts. In Figure~\ref{fig:profiles_nonbipartite}, Hollywood network seems to have a balanced distribution of all motifs, which may be also related to the fact that this network has the highest number of $P_4$'s as seen in Table~\ref{tab:datasets_all}. Figure~\ref{fig:profiles_bipartite} demonstrates that M1, M3, M6 and M11, the four motifs that contain a cycle on six vertices, are consistently among the least frequent across all four datasets. As expected, the counts fall as the motif gains edges. The five tree motifs ($M_{13}$--$M_{17}$) have the fewest edges and all rank in the upper half on every dataset, with $M_{13}$ the most frequent throughout; $M_1=K_{3,3}$, the only motif with nine edges, is the rarest everywhere.

    \section{Conclusion}
    We presented BiLight, a lightweight framework for estimating 6-vertex bipartite motif frequencies via weighted $P_4$ sampling, yielding unbiased and low-variance estimates. Our evaluation on large-scale networks shows that BiLight is accurate and scalable. The empirical comparison with Motivo~\cite{bressan2019motivoarxiv, bressan2019motivo, bressan2021faster} shows that our framework is substantially faster, and that the two agree closely on the datasets where Motivo's sampling covers all sixteen motifs.
    BiLight is tailored to the 17 bipartite 6-vertex motifs and can deliver all estimates quickly and consistently. We provide our source code that can as well be used towards generalizing our framework to non-bipartite motifs.

\onecolumn
    \appendices
    \clearpage
    \section{Runtime Result}
    \label{app:runtime}
\begin{figure}[H]
        \centering
        \includegraphics[width=0.75\linewidth]{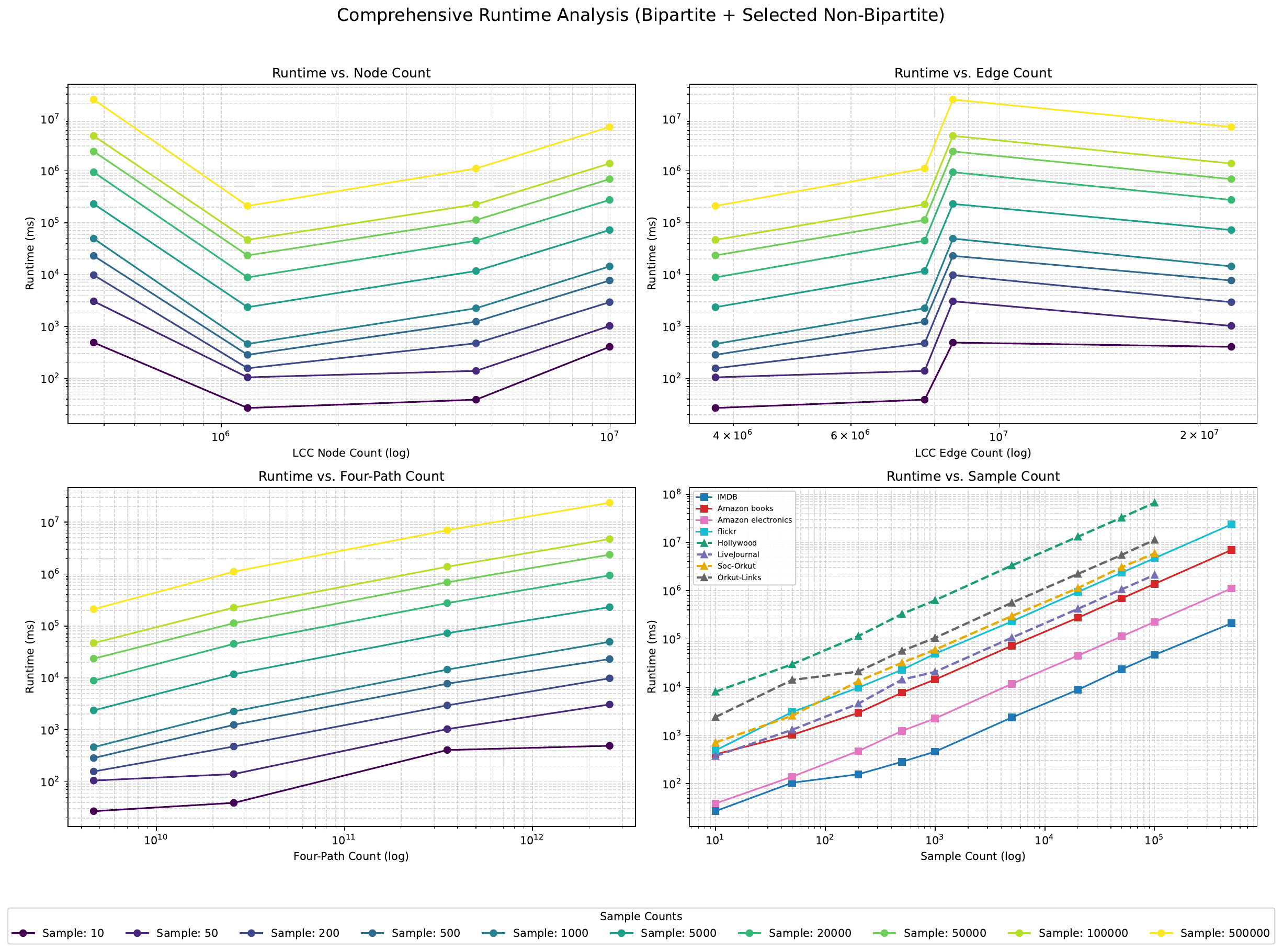}
        \caption{BiLight runtime analysis; each of the four panels plots estimation time (one-time setup excluded) against a different quantity. \emph{Top row (log scales):} runtime vs.\ largest connected component vertex count (left) and runtime vs.\ edge count (right). \emph{Bottom row:} runtime vs.\ $P_4$ count (left) and runtime vs.\ sample budget $k$ (right). The vertex-count, edge-count, and $P_4$-count panels use the bipartite datasets only; in the sample-budget panel, solid lines are the bipartite datasets and dashed lines are the four large non-bipartite graphs of Figure~\ref{fig:convergence_nonbipartite}.}
        \label{fig:runtime}
    \end{figure}

        Figure~\ref{fig:runtime} plots BiLight's estimation time; it grows approximately linearly with the sample budget $k$ on both bipartite and non-bipartite graphs.

    \section{Conversion Matrices}
    \label{app:conversion_matrices}
    Tables~\ref{tab:conversion_matrix_direct} and~\ref{tab:conversion_matrix} give $A$ and $A^{-1}$ for the 17 motifs (rows/columns M1--M17; blank cells are zero).

    \begin{table}[H]
        \centering
        \begin{minipage}[t]{0.49\linewidth}
        \centering
        \captionof{table}{Direct matrix $A$: $\mathbf{N} = A \mathbf{C}$. Entry $A_{i,j}$ is the number of copies of motif $M_i$ contained in motif $M_j$.}
        \label{tab:conversion_matrix_direct}
        \tiny
        \setlength{\tabcolsep}{0.5pt}
        \renewcommand{\arraystretch}{1.1}
        \rowcolors{2}{gray!12}{white}
        \begin{tabular}{@{}>{\raggedleft\arraybackslash}m{12pt}*{17}{!{\color{gray!35}\vrule}>{\raggedleft\arraybackslash}m{12pt}}@{}}
            \toprule
            & M1 & M2 & M3 & M4 & M5 & M6 & M7 & M8 & M9 & M10 & M11 & M12 & M13 & M14 & M15 & M16 & M17 \\
            \midrule
            M1  & 1 & & & & & & & & & & & & & & & & \\
            M2  & & 1 & & & & & & & & & & & & & & & \\
            M3  & 9 & & 1 & & & & & & & & & & & & & & \\
            M4  & & 8 & & 1 & & & & & & & & & & & & & \\
            M5  & 18 & & 4 & & 1 & & & & & & & & & & & & \\
            M6  & 18 & & 4 & & & 1 & & & & & & & & & & & \\
            M7  & & 12 & & 3 & & & 1 & & & & & & & & & & \\
            M8  & 36 & & 12 & & 4 & 2 & & 1 & & & & & & & & & \\
            M9  & & 12 & & 3 & & & & & 1 & & & & & & & & \\
            M10 & 36 & & 12 & & 2 & 4 & & & & 1 & & & & & & & \\
            M11 & 6 & & 2 & & & 1 & & & & & 1 & & & & & & \\
            M12 & & & & & & & & & & & & 1 & & & & & \\
            M13 & & 8 & & 3 & & & 2 & & & & & & 1 & & & & \\
            M14 & 9 & & 4 & & 2 & 1 & & 1 & & & & & & 1 & & & \\
            M15 & 36 & & 16 & & 6 & 6 & & 2 & & 2 & & & & & 1 & & \\
            M16 & & 24 & & 9 & & & 2 & & 4 & & & & & & & 1 & \\
            M17 & 36 & & 16 & & 4 & 8 & & 1 & & 2 & 6 & & & & & & 1 \\
            \bottomrule
        \end{tabular}
        \end{minipage}\hfill
        \begin{minipage}[t]{0.49\linewidth}
        \centering
        \captionof{table}{Inverse matrix $A^{-1}$: $\mathbf{C} = A^{-1} \mathbf{N}$. Row $i$, column $j$ correspond to motif $M_i$, $M_j$.}
        \label{tab:conversion_matrix}
        \tiny
        \setlength{\tabcolsep}{0.5pt}
        \renewcommand{\arraystretch}{1.1}
        \rowcolors{2}{gray!12}{white}
        \begin{tabular}{@{}>{\raggedleft\arraybackslash}m{12pt}*{17}{!{\color{gray!35}\vrule}>{\raggedleft\arraybackslash}m{12pt}}@{}}
            \toprule
            & M1 & M2 & M3 & M4 & M5 & M6 & M7 & M8 & M9 & M10 & M11 & M12 & M13 & M14 & M15 & M16 & M17 \\
            \midrule
            M1  & 1 & & & & & & & & & & & & & & & & \\
            M2  & & 1 & & & & & & & & & & & & & & & \\
            M3  & $-9$ & & 1 & & & & & & & & & & & & & & \\
            M4  & & $-8$ & & 1 & & & & & & & & & & & & & \\
            M5  & 18 & & $-4$ & & 1 & & & & & & & & & & & & \\
            M6  & 18 & & $-4$ & & & 1 & & & & & & & & & & & \\
            M7  & & 12 & & $-3$ & & & 1 & & & & & & & & & & \\
            M8  & $-36$ & & 12 & & $-4$ & $-2$ & & 1 & & & & & & & & & \\
            M9  & & 12 & & $-3$ & & & & & 1 & & & & & & & & \\
            M10 & $-36$ & & 12 & & $-2$ & $-4$ & & & & 1 & & & & & & & \\
            M11 & $-6$ & & 2 & & & $-1$ & & & & & 1 & & & & & & \\
            M12 & & & & & & & & & & & & 1 & & & & & \\
            M13 & & $-8$ & & 3 & & & $-2$ & & & & & & 1 & & & & \\
            M14 & 9 & & $-4$ & & 2 & 1 & & $-1$ & & & & & & 1 & & & \\
            M15 & 36 & & $-16$ & & 6 & 6 & & $-2$ & & $-2$ & & & & & 1 & & \\
            M16 & & $-24$ & & 9 & & & $-2$ & & $-4$ & & & & & & & 1 & \\
            M17 & 36 & & $-16$ & & 4 & 8 & & $-1$ & & $-2$ & $-6$ & & & & & & 1 \\
            \bottomrule
        \end{tabular}
        \end{minipage}
    \end{table}

    \onecolumn
    \section{Per-Motif Comparison}
    \label{app:permotif}

    \twocolumn
\ifCLASSOPTIONcompsoc
  \section*{Acknowledgments}
\else
  \section*{Acknowledgment}
\fi

The authors used Anthropic's Claude in preparing this work. The algorithms, the
local counting patterns, the theoretical analysis, the conversion matrices, and the
experiments are the authors' own. Claude's contributions lie mainly in
Sections~\ref{sec:preprocess} and~\ref{sec:samplingAlgo}: suggestion of Walker's alias
method in place of an $O(\log m)$ prefix-sum draw, correction of the preprocessing cost
to $O(\alpha m)$, brute-force verification to check the local counting patterns and
assistance with efficient presentation of experimental results. Claude's contributions to both
the text and the code were reviewed and revised by the authors, who take full
responsibility for the paper.

    \bibliographystyle{IEEEtran}
    \bibliography{sample_v3}

\end{document}